\documentclass[10pt,english,english,nofootinbib,notitlepage,superscriptaddress,aps,prd]{revtex4-1}

\usepackage[T1]{fontenc}
\usepackage{csquotes}
\usepackage{graphicx,color} 
\usepackage{xcolor}
\usepackage{rotating}
\usepackage{amssymb}
\usepackage{amsfonts}
\usepackage{amsmath}
\usepackage{xspace}
\usepackage{mathtools} 
\usepackage{verbatim}
\usepackage{breakcites}
\usepackage{accents}
\usepackage{dsfont}

\usepackage{listings}
\makeindex

\usepackage[thinlines]{easytable}

\usepackage[super]{nth}
\usepackage{braket}
\usepackage{cleveref}

\newcommand{\be}{\begin{equation}}
\newcommand{\ee}{\end{equation}}
\newcommand{\bea}{\begin{eqnarray}}
\newcommand{\eea}{\end{eqnarray}}
\newcommand{\id}{\mathds{1}}
\newcommand{\mompar}{\dot{\partial}}

\newcommand\ie{\mbox{\textit{i.\,e.}}\xspace}
\newcommand\cf{\mbox{c.\,f.}\xspace}
\newcommand\eg{\mbox{e.\,g.}\xspace}

\newcommand\D{\mathrm{d}}

\newcommand\op{\hat{\mathcal{O}}}
\newcommand\hil{\mathcal{H}}

\newcommand{\Mod}[1]{\ (\mathrm{mod}\ #1)}

\DeclareUnicodeCharacter{0301}{\'{e}} 

\begin{document}
%\preprint{}

\title{Towards quantum mechanics on the curved cotangent bundle}



%\thanks{}

\author{Fabian Wagner}
\email[]{fwagner@unisa.it}
\affiliation{Institute of Physics, University of Szczecin, Wielkopolska 15, 70-451 Szczecin, Poland\\
}
\affiliation{Dipartimento di Ingegneria Industriale, Universit\`a degli Studi di Salerno, Via Giovanni Paolo II, 132 I-84084 Fisciano (SA), Italy.}
\affiliation{INFN, Sezione di Napoli, Gruppo collegato di Salerno, Via Giovanni Paolo II, 132 I-84084 Fisciano (SA), Italy}
\date{\today}
\begin{abstract}
The minimal-length paradigm is a cornerstone of quantum gravity phenomenology. Recently, it has been demonstrated that minimal-length quantum mechanics can alternatively be described as an undeformed theory set on a nontrivial momentum space. However, there is no fully consistent formulation of these theories beyond Cartesian coordinates in ﬂat space and, in particular, no position representation. This paper is intended to take the ﬁrst steps in bridging this gap. We ﬁnd a natural position representation of the position and momentum operators on general curved cotangent bundles. In an expansion akin to Riemann normal coordinates with curvature in both position and momentum space, we apply the formalism perturbatively to the isotropic harmonic oscillator and the hydrogenic atom. Due to the symmetry of the harmonic oscillator under exchange of positions and momenta, we show that it is impossible to distinguish position- from momentum-space curvature with oscillators alone. Thus, we obtain an instantiation of Born reciprocity on the curved cotangent bundle, \ie in precisely the way Born originally envisioned. It manifests itself as a symmetry mixing UV and IR physics, reminiscent of T-duality in string theory.
\end{abstract}

%\pacs{}
%\keywords{}

\maketitle

\section{Introduction}
\label{intro}

Nontrivial momentum space, \ie a dependence of the length element on the direction of motion, has been investigated by mathematicians since it was first mentioned by Riemann in his habilitation dissertation on the workings of geometry \cite{Riemann54}. Significant contributions to the field were made by the likes of Finsler \cite{Finsler18} and Cartan \cite{Cartan34} and in the context of quantum groups by Drinfel'd \cite{Drinfeld88} and Majid \cite{Majid88,Majid90,Majid91}. These days, this approach spans the areas of Lagrangian and Hamiltonian geometries (for a detailed overview, consult \cite{Miron01,Miron12}).

Physicists, in contrast, have only gradually adopted the idea, for example, in the case of Wataghin to cure divergences through the nonlocal-field-theory program \cite{Wataghin37}. Born took a different avenue to arguing in favour of curvature in the cotangent bundle \cite{Born38,Born49}. As he noticed, quantum mechanics (as well as Hamiltonian mechanics) in flat space is symmetric under the exchange
\begin{equation}
    \hat{x}\rightarrow\hat{p},\hspace{1cm}\hat{p}\rightarrow -\hat{x}.\label{Bornrec}
\end{equation}
Owing to the term "reciprocal lattice" in condensed matter physics, this symmetry is nowadays called "Born reciprocity". It is explicitly broken in general relativity, where only position space is allowed to be curved. To retain the symmetry, it is necessary to consider momentum space to be nontrivial, hence, in Born's view \cite{Born38}, paving the way towards a unification of quantum theory and general relativity. 

The inclusion of nontrivial momentum space into physics was further developed mainly by Gol'fand \cite{Golfand59,Golfand62,Golfand63} and Tamm \cite{Tamm65,Tamm72}. This, in fact, lead to the canonical quantization of quantum field theories on curved momentum space \cite{Batalin89a,Batalin89b,Bars10}. Recent endeavours culminated in the Born geometry program \cite{Freidel13,Freidel14,Freidel15,Freidel17,Freidel18} in search of a construction capturing all mathematical structures behind Hamiltonian mechanics (symplectic), quantum theory (complex) and general relativity (metric) at once. Nowadays, curved momentum space is mostly found in applications to quantum gravity phenomenology, especially in deformed special (or general) relativity \cite{Kowalski-Glikman02c,Kowalski-Glikman03,Amelino-Camelia11,Carmona19,Relancio20a}. In the context of nonrelativistic quantum mechanics, this idea has already been investigated algebraically, \ie from the point of view of deformations of the Galilean algebra \cite{Papageorgiou10,Ballesteros21,Gubitosi21,Bosso:2023nst,Amelino-Camelia:2023rkg}. 

Since the early days of the conception of the theory \cite{Bronstein36}, the minimal-length concept has been a hallmark of quantum gravity \cite{Snyder46,Amati87,Gross87a,Gross87b,Amati88,Konishi89,Maggiore93b,Myung09b,Myung09c,Eune10}. Indeed, it suffices to combine heuristic arguments from general relativity and quantum mechanics to arrive at fundamental and absolute limitations to measurements \cite{Mead64,Mead66,Padmanabhan87,Ng93,Maggiore93a,Amelino-Camelia94,Garay94,Adler99a,Scardigli99,Capozziello99,Camacho02,Calmet04}. This effect can be modelled within nonrelativistic quantum mechanics through modifications to the algebra of observables \cite{Maggiore93c,Kempf94,Kempf96a,Benczik02,Das12}, resulting in generalized uncertainty principles (GUPs) (see \cite{Hossenfelder12} for a review and \cite{Bosso:2023aht} for a critical account of the state of the field). Similarly, there are analogous deformations leading to extended \cite{Bolen04,Park07,Bambi08a,Mignemi09,Ghosh09,CostaFilho16} and, a combination of the two, generalized extended uncertainty principles \cite{Bambi08a,Ong18}, in short EUPs and GEUPs, respectively. Consider, for instance, a modification of the form\footnote{Throughout this paper we set Planck's constant $\hbar$ equal to one.}
\begin{equation}
    \left[\hat{x},\hat{p}\right]=i\left(1+\wp^2\hat{x}^2+\ell^2\hat{p}^2\right)\label{eqn:GEUPHeis}
\end{equation}
with the length and momentum scales $\ell$ and $\wp,$ and the physical position and momentum $\hat{x}$ and $\hat{p},$ respectively.
According to the Robertson relation \cite{Robertson29} of position and momentum uncertainties $\Delta x$ and $\Delta p$, respectively, such a modified Heisenberg algebra implies the inequality
\begin{equation}
    \Delta x \Delta p\geq\frac{1}{2}\left(1+\wp^2\Delta x^2+\ell^2\Delta p^2\right).\label{GEuncrel}
\end{equation}
If $\ell\wp<1,$ this relation entails the existence of a minimal length $\Delta x\geq x_\text{min}=\ell/\sqrt{1-\ell^2\wp^2}$ as well as a minimal momentum $\Delta p\geq p_\text{min}=\wp/\sqrt{1-\ell^2\wp^2}$, \ie a maximal wave length $\lambda_{\text{max}}=1/p_\text{min}.$ Therefore, understanding the maximal wavelength as the radius of the cosmological horizon,\footnote{It is for this reason that the algebra in Eq. \eqref{GEuncrel} has a preferred position (the origin), where the position-dependent effect vanishes: The position of the cosmological horizon is observer-dependent. The observer, in turn, is situated in the origin.} it mimics quantum gravitational as well as horizon-induced effects. Despite a number of open issues (recently summarised in \cite{Bosso:2023aht}) such as the inverse soccer ball problem \cite{Quesne:2009vc,Tkachuk:2012gyq,Amelino-Camelia11}, the classical limit \cite{Casadio20}, depending on the relativistic completion, deformation \cite{Ali09,Bosso:2023nst} or straight breaking \cite{Lambiase17} of Lorentz invariance, this endeavour has lead to manifold phenomenological applications \cite{Bawaj14,Bushev19,Das11,Das08,Marin13,Marin14,Das08,Ali11b,Gao16,Khodadi18a,Petruzziello20,Buoninfante20}.

Recent times have seen a merger of the areas of the GUP and curved momentum space, first through indication \cite{Chang10,Schuermann18,Dabrowski19,Dabrowski20,Petruzziello21,Singh21,Wagner21b} and finally explicitly \cite{Wagner21a}. In particular, in \cite{Wagner21a,Wagner:2022dkc,Wagner:2022rjg} it was shown by direct construction that GUP-deformed theories can be mapped into quantum mechanics on nontrivial momentum space. Thus, in quantum theory a minimal length can be translated into a momentum-dependent metric. In other words, the dynamics may be described either by a deformed Heisenberg algebra of the type showcased in Eq. \eqref{eqn:GEUPHeis} with (apparent) flat-space Hamiltonian or by conjugate phase-space variables (satisfying the ordinary Heisenberg algebra) with a Hamiltonian describing motion on the curved cotangent bundle. As a result, the curvature in momentum space corresponding to the usually applied quadratic GUP is directly proportional to the noncommutativity of the coordinates in position space (see also \cite{Amelino-Camelia11}), a fact, which continues to hold for the nonanalytic linear quadratic GUP \cite{Wagner:2022dkc}.

In other words, Born reciprocity on nontrivial backgrounds and the GEUP are two sides of the same coin. This analogy is already indicated by the symmetry of Eq. \eqref{GEuncrel} under the transformation \cite{Bambi08a,Dabrowski19}
\begin{equation}
   \wp\Delta x\longleftrightarrow\wp \Delta p,
\end{equation}
which shows clear reminiscence of Eq. \eqref{Bornrec} (note that by definition $\Delta\hat{\mathcal{O}}=\Delta(-\hat{\mathcal{O}})$ for all operators $O$). Therefore, a consistent geometric formulation of quantum mechanics on the curved cotangent bundle would permit a full description of minimal-length and minimal-momentum quantum mechanics, a subject that was recently at the heart of debates surrounding the possible disappearance of the Chandrasekhar limit in the presence of a minimal length \cite{Rashidi15,Ong18,Du22}.

While the conventional approach to quantum mechanics on curved backgrounds pioneered by Bryce deWitt's approach \cite{DeWitt52} proves successful in the regime of curved position space in the position representation and curved momentum space in the momentum representation, it cannot be applied once the line element is of the form\footnote{Given the inverse metric $g^{ij}(x,p)$, we can also define a line element in momentum space of the form $g^{ij}\delta p_i\delta p_j.$ Here, the differential $\delta p_i$ is defined in Sec. \ref{sec:curvcot}.}
\begin{align}
	\D s^2=&g_{ij}\left(x,p\right)\D x^i\D x^j.\label{lengthel}
\end{align}
Thus, there is a strong motivation for the construction of a quantum mechanical formalism which maintains consistency on the curved cotangent bundle. The present paper delineates an avenue towards the quantum description of such a theory of a single particle in the position representation.

In that vein, we choose to generalize deWitt's ideas \cite{DeWitt52} in the most minimal way while always maintaining covariance with respect to spatial coordinate transformations in accordance with the theory of generalized Hamilton geometries \cite{Miron01,Miron12}. Correspondingly, we promote the measure of the Hilbert space, \ie the volume form, to an Hermitian operator. Furthermore, we split it into two pieces to merge them symmetrically with each wave function in the scalar product, turning them into scalar densities.\footnote{Of course, we are not the first to propose to absorb the measure into the wave function. Indeed, this representation is integral to geometric quantisation \cite{Woodhouse97}, and has been proposed as a remedy to the malaise of single-particle quantum mechanics on evolving background geometries \cite{Mostafazadeh:2018dpg,Schwartz:2018pnh}.} Then, assuming that the scalar product of their eigenvectors consists of plane waves, the momentum operator possesses a particularly simple position representation. Furthermore, we find representations for the Hamiltonian of the free particle and, equivalently, the geodesic distance from the origin. From the latter, we can construct interaction potentials. Finally, we turn to phenomenology dealing with central potentials, particularly the isotropic harmonic oscillator and the hydrogenic atom, on a background harbouring curvature in position as well as momentum space described by an expansion akin to Riemann normal coordinates. In this context, we find that the isotropic harmonic oscillator retains its Born reciprocal nature also on the curved cotangent bundle. In particular, this behaviour is indicated by a symmetry akin to T-duality in string theory \cite{Sathiapalan86,Alvarez94}. Thus, combining the concepts of Born reciprocity and the generally curved cotangent bundle, we obtain the kind of UV-IR mixing which is expected from some approaches to quantum gravity \cite{Minwalla99,Freidel:2021wpl}.

Note that this article concerns quantum dynamics on classical backgrounds which happen to be generalized Hamilton geometries -- not quantum gravity. Generalized Hamilton geometries are examples of quantum spaces, phenomenologically incorporating possible nonperturbative quantum-gravity effects. Unlike deWitt \cite{DeWitt52}, we do not apply our results to the quantisation of the gravitational field to obtain possible modifications of quantum geometrodynamics. A quantisation of Hamilton geometries, allowing for both dynamical position space and dynamical momentum space simultaneously, would have to start from an action principle on these spaces. Such an action principle has been proposed in \cite{Pfeifer:2011xi}.\footnote{Strictly speaking, this action principle governs the dynamics of Finsler geometries. These provide the Lagrangian description of Hamilton geometries.} While a generalisation of quantum geometrodynamics along these lines would be a worthwhile, if demanding pursuit, it has not been accomplished yet, and goes beyond the scope of the present paper.

The paper is organized as follows. We briefly explain some basics of the theory of generalized Hamilton spaces in Sec. \ref{sec:curvcot}. In Sec. \ref{sec:dW} we review the conventional approach to quantum mechanics in curved space. The main formalism is introduced as a generalization of the customary formalism in Sec. \ref{sec:formalism}. Sec. \ref{sec:application} is devoted to phenomenology, yielding a perturbative application of the approach to central potentials. Finally, we summarize the results and conclude in Sec. \ref{sec:concl}.

\section{The curved cotangent bundle\label{sec:curvcot}}

This section is intended to briefly summarize the ingredients of Hamilton geometries required for the purpose of the present paper, and introduce the notation. For a comprehensive account consult \cite{Miron01,Miron12}. In generalized Hamilton geometries the distance between two points can be momentum-dependent in accordance with Eq. \eqref{lengthel}.  Yet, the formalism retains covariance with respect to diffeomorphisms
\begin{equation}
    x^i\rightarrow X^i(x),\hspace{1cm}p_i\rightarrow P_i=\frac{\partial x^j}{\partial X^i}p_j.\label{gencoordtrans}
\end{equation} 
If the cotangent bundle exhibits curvature -- even if solely position space is curved --, the canonical variables do not provide a clear partition into position or momentum space any more. In particular, a nonlinear connection $N_{ij}(x,p)$ is required as a bookkeeping device dividing the cotangent bundle into the horizontal (position) and vertical (momentum) distributions. Under reasonable conditions, the nonlinear connection can be derived from the metric in a canonical way. For example, in the Riemannian case, \ie for purely positional curvature, its canonical version reads
\begin{equation}
    N_{ij}=p_k\Gamma^k_{ij}(x),\label{levicivnonlin}
\end{equation}
with the Christoffel symbols of the Levi-Civita connection $\Gamma^k_{ij}.$

Given a nonlinear connection, we can define the orthogonal vector fields on phase space
\begin{equation}
    \frac{\delta}{\delta x^i}=\frac{\partial}{\partial x^i}+N_{ij}\frac{\partial}{\partial p_j},\hspace{1cm}\frac{\partial}{\partial p_i},\label{modder}
\end{equation}
both of which transform covariantly under the coordinate transformation \eqref{gencoordtrans}
\begin{equation}
    \frac{\delta}{\delta x^i}=\frac{\partial X^j}{\partial x_i}\frac{\delta}{\delta X^j},\hspace{1cm}\frac{\partial}{\partial p_i}=\frac{\partial x^i}{\partial X^j}\frac{\partial}{\partial P_j}.
\end{equation}
These two vector fields are related by a map $F(\delta_i)=-g_{ij}\dot{\partial}^j$ (where the metric $g_{ij}$ has been defined in Eq. \eqref{lengthel}), $F(\dot{\partial}^i)=g^{ij}\delta_j,$ with $\delta_i\equiv\delta/\delta x^i,$ $\dot{\partial}^i\equiv\partial/\partial p_i.$ Thus, the present system admits an almost complex structure ($F^2=-\id$), which is precisely the mathematical premise underlying quantum mechanics -- in physicist's terms: Born reciprocity.

Crucially, the positional partial derivative is modified, yielding a curvature tensor on the cotangent bundle
\begin{equation}
    \left[\frac{\delta}{\delta x^i},\frac{\delta}{\delta x^j}\right]\equiv R_{kij}\frac{\partial}{\partial p_k},
\end{equation}
which for Riemannian backgrounds assumes the form
\begin{equation}
    R_{kij}=p_mR^{m}_{~kij}(x),
\end{equation}
with the Riemann curvature tensor $R^{m}_{~kij}.$ Furthermore, the antisymmetric part of the nonlinear connection yields the torsion on the cotangent bundle which is assumed to vanish for the remainder of the present paper.

Similarly, there is a basis of one-forms
\begin{equation}
    \D x^i,\hspace{1cm}\delta p_i=\D p_i-N_{ji}\D x^j,\label{fundforms}
\end{equation}
which, too, transform as expected
\begin{equation}
    \D x^i=\frac{\partial x^i}{\partial X^j}\D X^j,\hspace{1cm}\delta p_i=\frac{\partial X^j}{\partial x^i}\delta p_j.
\end{equation}
Here, the symbol $\delta$ is to be understood as an exterior covariant derivative. If the space under consideration is endowed with a line element as in Eq. \eqref{lengthel}, we can construct the covariant integration measures
\begin{equation}
    \sqrt{g}\D^dx,\hspace{1cm} \frac{1}{\sqrt{g}}\delta^d p,\label{modint}
\end{equation}
with the determinant of the metric $g=\det g_{ij},$ and where $d$ counts the number of dimensions.
 
From Eqs. \eqref{modder} and \eqref{modint}, it is clear, that it is much easier to maintain a position representation involving only position-dependent wave functions $\psi(x)$ (such that $\delta_i\psi=\partial_i\psi$) than to deal with the momentum representation. Therefore, unless $N_{ij}=0,$ \eg for purely momentum-dependent metrics, we shall restrict ourselves to the former case. %The inclusion of momentum-dependence will likely require a doubling of the phase space variables as it is done in Born geometries \cite{Freidel13,Freidel14,Freidel15,Freidel17,Freidel18} and double field theory \cite{Siegel93a,Siegel93b,Hull09}.

Incidentally, this simplification in position space is also why it is possible to deal with quantum mechanics on Riemannian backgrounds in the way described in the following section, \ie without reverting to the nonlinear connection.

\section{Conventional approach to quantum mechanics on curved spaces}\label{sec:dW}

The problem of a nonrelativistic quantum particle on a curved spatial background manifold, governed by the classical Hamiltonian $H=g^{ij}(x)p_ip_j/2M+V(x),$ was first considered by Bryce deWitt \cite{DeWitt52}, based on the requirement of maintaining covariance with respect to arbitrary coordinate transformations in every step. Since its conception it has been generalized to general background group manifolds \cite{Dowker70,Lindesay13}, supersymmetric contexts \cite{deAlfaro87a,deAlfaro87b} and formulations describing particles constrained to curved surfaces \cite{Homma90,Ikegami90}. Modern applications include \eg  \cite{Dorn10,Bravo15,Petruzziello21,Wagner21b}.

In general,  curved $d$-dimensional backgrounds are most intuitively introduced into quantum mechanics by virtue of a covariant measure of the Hilbert-space scalar product 
\begin{equation}
\braket{\psi|\phi}\equiv\int\D^d x \sqrt{g(x)}\psi^*\phi,\label{scalarprod1}
\end{equation}
where $g$ denotes the determinant of the background metric. Then, the wave functions as well as the scalar product transform as scalars under coordinate transformations. This can be traced back to the definition of the position operator as
\begin{equation}
\hat{x}^i\equiv\int \D^d x\sqrt{g} x^i\ket{x}\bra{x},\label{xop1}
\end{equation}  
with its eigenstates $\ket{x}$ and eigenvalues $x^i.$ The assumption that $\hat{x}^i$ is Hermitian implies the orthonormality of its eigenstates
\begin{align}
\mathds{1}=						&\int \D^d x\sqrt{g}\ket{x}\bra{x},\label{complete1x}\\
\braket{x|x'}=	&\frac{\delta^d(x-x')}{\sqrt{g}},\label{ortho1x}
\end{align}
with Dirac's $\delta$-distribution in $d$ dimensions $\delta^d(x-x').$ Then, the completeness relation \eqref{complete1x} can be used to express every state $\ket{\psi}\in\hil$ in the position representation as
\begin{equation}
\ket{\psi}=\int\D^d x\sqrt{g}\braket{x|\psi}\ket{x}\equiv\int\D^d x\sqrt{g}\psi(x)\ket{x}.
\end{equation}
Application of Eq. \eqref{ortho1x} then immediately leads to the scalar product \eqref{scalarprod1}. 

Assuming that the position and momentum operators continue to obey the Heisenberg algebra
\begin{equation}
    [\hat{x}^i,\hat{p}_j]=i\delta^i_j,\hspace{.7cm}[\hat{x}^i,\hat{x}^j]=[\hat{p}_i,\hat{p}_j]=0,\label{HeisAlg}
\end{equation}
the momentum operator has to be modified in order to be Hermitian with respect to the invariant measure. Thus,  we obtain the position representation \cite{DeWitt52}
\begin{equation}
\hat{p}_i\psi=-i\left(\partial_i+\frac{1}{2}\Gamma^j_{ij}\right)\psi=\frac{1}{\sqrt[4]{g}}\partial_i\left(\sqrt[4]{g}\psi\right),\label{mom1}
\end{equation}
with the Christoffel symbol $\Gamma^k_{ij}.$ As a matter of fact, this corresponds to a covariant derivative acting on scalar density of weight $1/2$ as necessitated by the determinant of the metric appearing in the covariant volume measure.

Diffeomorphisms (in de Witt's terminology point transformations) as provided in Eq. \eqref{gencoordtrans}, then, have the quantum analogue
\begin{align}
	\hat{x}^i\to\hat{X}^i,&&\hat{p}_i\to\hat{P}_i=\left\{\frac{\D \hat{x}^j}{\D\hat{X}^i}\hat{p}_j\right\},\label{eqn:QuantDiff}
\end{align}
as can be derived from Eq. \eqref{mom1}, using solely the transformation properties of the partial derivative and the determinant of the metric.

The quantization of the Hamiltonian of a free particle, classically given as $H_{\text{fp}}=g^{ij}p_ip_j/2M,$ with the mass of the particle in question $M,$ is naturally object to operator-ordering ambiguities. In general, it can be represented in manifestly covariant fashion as
\begin{equation}
    \hat{H}_{\text{fp}}\psi=-\frac{1}{2M}\left[\frac{1}{\sqrt{g}}\partial_i\left(\sqrt{g}g^{ij}\partial_j \psi\right)+a R\left(\hat{x}\right)\psi\right],
\end{equation}
where the dimensionless parameter $a$ quantifies the ambiguity and we introduced the scalar curvature $R(x).$ This ambiguity can be resolved by applying geometric calculus (see  \cite{Pavsic01} and references therein),\footnote{The idea underlying this reasoning lies in defining the momentum operator as a scalar Euclidean Dirac operator $\gamma^i(x) \partial_i,$ with the matrices $\gamma^i$ satisfying $\{\gamma^i,\gamma^j\}=2g^{ij}.$ This is how the use of vector operators, whose expectation values are coordinate dependent in single-particle quantum mechanics, can be avoided. Then, the Hamiltonian is to be defined as $\hat{H}\psi= \gamma^i\partial_i(\gamma^j\partial_j\psi)/2M=\Delta\psi/2M,$ implying that $a=0.$} fixing the parameter $a=0,$ \ie leading to the representation
\begin{equation}
    \hat{H}_{\text{fp}}\psi=-\frac{1}{2M}\frac{1}{\sqrt{g}}\partial_i\left(\sqrt{g}g^{ij}\partial_j \psi\right)\equiv-\frac{1}{2M}\Delta\psi.\label{hfpdW}
\end{equation}
The last equality defines the Laplace-Beltrami operator, which is clearly invariant under diffeomorphisms as can be derived from the transformation properties of partial derivatives, the metric and its determinant. 

Summarizing, the quantization of a general Hamiltonian on a curved background manifold in the position representation reads
\begin{equation}
    \hat{H}\psi=\left[-\frac{1}{2M}\Delta+V(x)\right]\psi.\label{eqn:deWittHam}
\end{equation}
Here, the potential $V(x)$ is comprised of interactions with very massive objects (scalar background fields) such that backreaction can be neglected.\footnote{Note that interactions have to depend on positions solely through the geodesic distance between two particles to maintain covariance.} The resulting quantum formalism can be constructed by analogy with the textbook treatment, bearing in mind that it is restricted to the position representation. Indeed, a momentum-space equivalent of Eq. \eqref{eqn:deWittHam} does, as of yet, not exist.

As long as the metric depends solely on momenta, this ansatz can, by Born reciprocity, be equivalently applied to nontrivial momentum space in the momentum representation. But, analogously, it cannot yield a corresponding description in position space. Thus,  in order to treat general metrics describing the curved cotangent bundle, the conventional approach has to be generalised. 

\section{Quantizing on the curved cotangent bundle}\label{sec:formalism}

Once it depends on positions and momenta, the determinant of the metric appearing in the volume form in the scalar product has to be promoted to an operator $\sqrt{\hat{g}}\D^d x.$ Formally, we can construct the operator version of Eq. \eqref{scalarprod1},
\begin{equation}
	\braket{\psi|\phi}_{\sqrt{\hat{g}}}=\int\psi(x)^*\sqrt{\hat{g}(\hat{x},\hat{p})}\psi(x),\label{eqn:GenInProd}
\end{equation}
where we introduced some continuous set of states $\ket{x}$ such that $\psi(x)=\braket{x|\psi}_\id$ is a square-integrable function, and follow the notation of \cite{Anderson:1992lns,Anderson:1993im} indicating the measure as a subscript whenever it is operator-valued. Using the same notion, we can rewrite Eq. \eqref{eqn:GenInProd} as
\begin{equation}
	\braket{\psi|\phi}_{\sqrt{\hat{g}}}=\braket{\psi|\sqrt{\hat{g}}\phi}_{\id}.
\end{equation}
In other words, the operator $\sqrt{\hat{g}}$ acts on the space of square integrable functions in $d$ dimensions with unit measure $L^2(\mathbb{R}^d,\D^d x).$ For the bilinear introduced in Eq. \eqref{eqn:GenInProd} to be an inner product of a Hilbert space, it has to satisfy
\begin{align}
	\braket{\psi|\psi}_{\sqrt{\hat{g}}}\in \mathbb{R}_+.\label{eqn:HilbertCond}
\end{align}
This is the case if and only if $\hat{g}$ has solely real, positive (non-zero) eigenvalues, \ie it is Hermitian and positive definite with respect to the unit measure. In other words, $g(x,p)$ has to be a positive, analytic function of its arguments. Therefore, throughout this paper, we assume the determinant of the metric to be non-degenerate, excluding singular spaces. If the condition \eqref{eqn:HilbertCond} is satisfied, we can normalise the considered states such that
\begin{equation}
	\braket{\psi|\phi}_{\sqrt{\hat{g}}}=1
\end{equation}
as in ordinary quantum theory.

In order to construct a position representation in a symmetric way, denote the square root of the operator-valued measure as $\hat{\mu}(\hat{x},\hat{p})\equiv\sqrt[4]{\hat{g}}.$ This operator necessarily possesses solely real, positive eigenvalues just as $\hat{g}$ does. It is understood as an input to the formalism including a given (necessarily symmetric) ordering if there are ordering ambiguities. Then, the position operator can be defined as (\cf Eq. \eqref{xop1})
\begin{equation}
\hat{x}^i\equiv\int \D^d x x^i\ket{\hat{\mu}x}\bra{\hat{\mu}x}.\label{posop}
\end{equation}
Thus, its eigenstates are of the form $\ket{\hat{\mu}x}=\hat{\mu}\ket{x}$ such that $\hat{x}^i\ket{\hat{\mu}x}=x^i\ket{\hat{\mu}x}.$ Similarly to the conventional approach, these eigenstates furnish an orthonormal basis, hence satisfying
\begin{align}
\int \D^d x\ket{\hat{\mu}x}\bra{\hat{\mu}x}=&\mathds{1},\label{completex}\\
\braket{\hat{\mu}x|\hat{\mu}x'}_\id=	&\delta^d(x-x'),\label{orthox}
\end{align}
which generalize Eqs. \eqref{complete1x} and \eqref{ortho1x}. Then, every state in the Hilbert space can be expanded in the given basis, yielding
\begin{equation}
    \ket{\psi}=\int\D^d x\braket{\hat{\mu}x|\psi}_\id\ket{\hat{\mu} x}.
\end{equation}
Defining the position space wave function as $\psi\equiv\braket{x|\psi},$ we can express amplitudes in terms of the density-valued wave function\footnote{The quantities $\Phi,\Psi\dots$ and their correspondents in momentum space $\tilde{\Phi},\tilde{\Psi}\dots$ change under coordinate transformations not as scalar functions but as scalar densities of weight $1/2$ and $-1/2,$ respectively.} $\Psi\equiv\braket{\hat{\mu}x|\psi}=\hat{\mu}(\psi),$ where the action of the measure on wave functions $\hat{\mu}(\psi)$ depends on the exact form of $\hat{\mu}.$ This leads to the corresponding scalar product
\begin{align}
\braket{\psi|\phi}_{\sqrt{\hat{g}}}=	&\int\D^d x\braket{\psi|\hat{\mu}x}_\id\braket{\hat{\mu}x|\phi}_\id,\\
			\equiv	&\int\D^d x \Psi^*\Phi,\\
			=&\braket{\Psi|\Phi}_\id.\label{scapro}
\end{align}
Note that if the measure is a function of the position operator only, \ie $\hat{\mu}=\hat{\mu}(\hat{x}),$ it commutes with the position operator, thus rendering both of them simultaneously diagonalisable.  Then, $\ket{x}$ can be understood as eigenstate of $\hat{\mu},$ which implies $\ket{\hat{\mu}x}=\mu(x)\ket{x},$ recovering the description \`a la deWitt (\cf Eqs. \eqref{xop1}, \eqref{complete1x} and \eqref{ortho1x}). We stress, however, that the formalism introduced in the present section also applies to the case $[\hat{\mu},\hat{x}^i]\neq0.$ As can be inferred from  Eq. \eqref{scapro}, this procedure provides a map from quantum mechanics on the curved cotangent bundle into ordinary flat space-quantum mechanics, given the actions of operators on density-valued wave functions like $\Phi$ and $\Psi.$ 

\subsection{A note on the momentum representation}

As the framework developed in this paper is intended to be invariant under coordinate transformations, the discussion in momentum space is highly nontrivial. In particular, the measure $\D^dp/\sqrt{g}$ does not reflect this principle. Instead, equation \eqref{fundforms} indicates that the unique covariantly transforming basis of one-forms in momentum space is
\begin{equation}
    \omega_i\equiv \delta p_i. 
\end{equation}
However, this set of forms is not closed, \ie $\D \omega_i\neq 0,$ unless $R_{ijk}=\dot{\partial}^kN_{ij}=0,$ which is already violated by a position-dependent metric (\cf Eq. \eqref{levicivnonlin}). Thus, they cannot be exact, a necessary condition for them to be derivable from some new kind of phase space coordinate. Therefore, an application of the reasoning introduced above to momentum space is not immediate. Instead, it seems plausible that a full treatment will likely require a doubling of the phase space coordinates as in Born geometries \cite{Freidel13,Freidel18} or double field theory \cite{Siegel93a,Siegel93b,Hull09}. This line of research will be dealt with in future work, while we presently restrict ourselves mainly to the position representation.

There is an exception to this rule, however. If the metric turns out to be purely momentum-dependent, \ie if $g_{ij}=g_{ij}(p)\neq g_{ij}(x,p),$ the nonlinear connection can be chosen to vanish \cite{Miron01,Wagner21a,Relancio21}. By analogy with Eq. \eqref{posop}, we can then formally define the momentum operator as
\begin{equation}
\hat{p}_i=\int \D^d p p_i\ket{\hat{\mu}^{-1}p}\bra{\hat{\mu}^{-1}p}.\label{momop}
\end{equation}
In order to give meaning to this equation, we have to properly define the inverse of $\hat{\mu}.$ If $\hat{\mu}=\hat{\mu}(\hat{p})\neq\hat{\mu}(\hat{x},\hat{p}),$ there are no ordering ambiguities. Furthermore, the positiveness and analyticity of $g$ allow to express $\hat{\mu}$ in a Taylor series expansion
\begin{equation}
	\hat{\mu}(\hat{p})=\sum_{n=0}^\infty \bar{\mu}^{i_1\dots i_n}\hat{p}_{i_1}\dots\hat{p}_{i_n},
\end{equation}
where we introduced the c-number-valued series coefficients $\bar{\mu}^{i_1\dots i_n}.$ Then, the inverse operator $\hat{\mu}^{-1}$ can be defined recursively by virtue of the Lagrange inversion theorem, and will be both a left and a right invariant.

The eigenstates of the momentum operator $\ket{\hat{\mu}^{-1}p}$ obey the relations
\begin{align}
\int \D^d p\ket{\hat{\mu}^{-1}p}\bra{\hat{\mu}^{-1}p}=						&\mathds{1},\label{completep}\\
\braket{\hat{\mu}^{-1}p|\hat{\mu}^{-1}p'}_\id=	&\delta^d(p-p').\label{orthop}
\end{align}
Accordingly, this provides us with the momentum representation
\begin{equation}
    \ket{\psi}=\int\D^d p\braket{\hat{\mu}^{-1}p|\psi}_\id\ket{\hat{\mu}^{-1} p}\equiv\int\D^d p\hat{\mu}^{-1}(\tilde{\psi})(p)\ket{\hat{\mu}p},
\end{equation}
where $\tilde{\psi}(p)\equiv\braket{p|\psi}.$ Defining $\tilde{\Psi}(p)\equiv\braket{\hat{\mu}^{-1}p|\psi}_\id=\hat{\mu}^{-1}(\tilde{\psi}),$ we obtain the scalar product
\begin{align}
\braket{\psi|\phi}=	&\int\D^d p\braket{\psi|\hat{\mu}^{-1}p}_\id\braket{\hat{\mu}^{-1}p|\phi}_\id,\\
						 \equiv	&\int\D^d p \tilde{\Psi}^*\tilde{\Phi}.
\end{align}
Before this formalism can be applied to examples, though, it remains to be shown that this construction is well-defined.

\subsection{Representations of conjugate operators}

The phase space basis displayed in Eq. \eqref{modder} indicates that the momentum operator ought to be represented in position space as
\begin{equation}
    \Braket{\hat{\mu} x|\hat{p}_i\psi}=-i\delta_i\Psi(x)=-i\partial_i\Psi(x).
\end{equation}
Note that this last equality does not imply that the nonlinear connection vanishes, but follows from the fact that wave functions solely depend on positions. In particular, the corresponding commutator with the position operator reads
\begin{equation}
    \left[x^i,-i\delta_j\right]\Psi=i\left(\delta_j^i\Psi-\dot{\partial}^iN_{jk}\dot{\partial}^k\Psi\right)=i\delta ^i_j\Psi.
\end{equation}
Thus, the position and momentum operators continue to satisfy the Heisenberg algebra \eqref{HeisAlg}. Furthermore, a general amplitude featuring the momentum operator reads
\begin{equation}
    \braket{\psi|\hat{p}_i\phi}_\id=\int\D^d x\Psi^*\left(-i\partial_i\Phi\right),
\end{equation}
which is trivially symmetric and in the case of a purely position-dependent measure just recovers Eq. \eqref{mom1}. In generalized Hamilton geometries, the metric and, consequently, its determinant, the partial derivative and the invariant volume measure all retain the transformation properties (under diffeomorphisms) they assume in Riemannian geometry. Consequently there continues to be a quantum analogue of the diffeomorphisms as given in Eq. \eqref{eqn:QuantDiff}, having the same effect on the proposed ansatz as in de Witt's formalism.

Analogously, if the metric is purely momentum-dependent, we obtain for the momentum representation of the position operator
\begin{equation}
        \braket{\hat{\mu}^{-1}p|\hat{x}^i\psi}_\id=\hat{x}^i\tilde{\Psi}=i\dot{\partial}^i\tilde{\Psi},
\end{equation}
in accordance with Eq. \eqref{modder}. Then, the Fourier transform can be generalized such that it reads
\begin{align}
    \tilde{\Psi}=\frac{1}{\sqrt{2\pi}}\int\D^d x \Psi e^{-i p_i x^i},&&\Psi=\frac{1}{\sqrt{2\pi}}\int\D^d p \tilde{\Psi}e^{ip_ix^i}.
\end{align}
Thus, it is at least possible to find a complete picture of quantum mechanics on curved momentum space, dual to the GUP-formalism.

It is evident that the representations of the position and momentum operators defined in this section are indeed Hermitian (due to the symmetric choices in Eqs. \eqref{posop} and \eqref{momop}) and satisfy the canonical commutation relations \eqref{HeisAlg}. Thus, $\hat{\mu},$ itself a real function of $\hat{x}^i$ and $\hat{p}_j,$ can be made Hermitian by suitable symmetrization and is well-defined modulo operator ordering ambiguities. We conclude that the procedure outlined above is consistent.

\subsection{Free particle}

The Hamiltonian of a free particle of mass $M$ subject to a general metric $g^{ij}(x,p)$ is not expected to retain its ordinary form. In particular, it measures the distance from the origin in momentum space. If momentum space is curved, however, the only covariant quantity at hand is the geodesic distance $\sigma_p,$ which, at the classical level, satisfies the differential equation
\begin{equation}
	g_{ij}\mompar^i(\sigma_p^2)\mompar^j(\sigma_p^2)=4\sigma_p^2.
\end{equation}
Therefore, we can generally express the free-particle Hamiltonian as \cite{Franchino-Vinas:2022fkh,Franchino-Vinas:2023rcc}
\begin{equation}
	\hat{H}=\frac{\sigma_p^2}{2M}=\frac{g_{ij}\tilde{p}^i\tilde{p}^j}{2M},
\end{equation}
where we introduced the vector field $\tilde{p}^i(x,p)=\mompar^i (\sigma_p^2).$ 

Recall that the Hamiltonian operator on a curved position space is given by the Laplace-Beltrami operator, \ie Eq. \eqref{hfpdW}. Thus, its general matrix element reads
\begin{equation}
    \braket{\psi|\hat{H}_{\text{fp}}\phi}=-\frac{1}{
2M}\int \D^d x\psi^*\partial_i\left(\sqrt{g}g^{ij}\partial_j\phi\right).\label{kinenamp1}
\end{equation}
In the more general case treated here the representation of the Hamiltonian can be naturally generalized to the form
\begin{equation}
	\hat{H}_{\text{fp}}\Psi=\frac{1}{4M}\hat{\mu}^{-1}\tilde{p}_i(\hat{x},\hat{p})\{\hat{\mu}^2,\hat{g}^{ij}\}\tilde{p}_j(\hat{x},\hat{p})\hat{\mu}^{-1}\Psi,\label{eqn:DefKinEnCurvCot}
\end{equation}
where the metric and the vector $\tilde{p}_i$ have been promoted to Hermitian operators, applying Weyl-ordering. In order to give meaning to the inverse of $\hat{\mu}$ now that it depends on noncommuting variables ($\hat{x}$ and $\hat{p}$), we again note that $g$ is assumed to be a positive, analytic function of both of its arguments, thus allowing for a Taylor series expansion
	\begin{equation}
	\hat{\mu}=\sum_{n,m=0}^\infty\sum_{\mathcal{O}}\accentset{\mathcal{O}}{\mu}_{i_1\dots i_n}^{j_1\dots j_m}\hat{\mathcal{O}}\left( \hat{x}^{i_1}\dots\hat{x}^{i_n}\hat{p}_{j_1}\dots\hat{p}_{j_n}\right),
\end{equation}
where $\mathcal{O}$ enumerates all possible symmetric orderings which are conceivable at the given order of expansion, and the operator $\hat{O}$ implements these orderings, and we introduced the real c-number-valued series coefficients $\accentset{\mathcal{O}}{\mu}_{i_1\dots i_n}^{j_1\dots j_m}$.
Given such a series expansion, it is possible to use a generalisation of the Lagrange inversion theorem to noncommuting variables \cite{Gessel:1980abf}, to obtain the series coefficients of the inverse series.

Furthermore, in Eq. \eqref{eqn:DefKinEnCurvCot} the metric does not necessarily commute with the square root of its determinant $\hat{\mu}^2.$ Therefore, we introduced the anti-commutator $\{\hat{A},\hat{B}\}=\hat{A}\hat{B}+\hat{B}\hat{A}$ (for operators $A$ and $B$) to maintain symmetry. As a symmetric product of Hermitian operators, the free-particle Hamiltonian is manifestly Hermitian. Besides, the expression $\hat{H}_{\text{fp}}\Psi$ transforms under coordinate transformations as a scalar density of weight $1/2,$ as it should.

A general amplitude featuring the Hamiltonian of a free particle can thus be expressed as
\begin{equation}
    \braket{\psi|\hat{H}_{\text{fp}}\phi}_{\sqrt{\hat{g}}}=-\frac{1}{4M}\int \D^d x\Psi^*\hat{\mu}^{-1}\tilde{p}^i(x,\partial)\left[\{\hat{\mu}^2,\hat{g}^{ij}\}\tilde{p}^j(x,\partial)\left(\hat{\mu}^{-1}\Phi\right)\right],
\end{equation}
which in the case of a purely position-dependent metric exactly recovers Eq. \eqref{kinenamp1}. As the momenta, the invariant volume measure, the metric and its determinant transform under diffeomorphisms exactly as in de Witt's formalism, this amplitude is a scalar. Thus, the quantum analogue to diffeomorphisms introduced in \cite{DeWitt52} persists in the present formalism.

Having derived the action of the basic operators in the given framework, we can now turn to phenomenology.

\section{Phenomenology}\label{sec:application}

Gravitational corrections, both quantum and classical, to quantum mechanical experiments are usually negligibly small. For instance, the relative correction of the earth's gravitational field to the energy spectrum of the hydrogen atom is of the order $10^{-38}$ \cite{Zhao07}, while the corresponding magnitude stemming from a Planckian GUP is expected to lie around $10^{-44}$ \cite{Bouaziz10,AntonacciOakes13,Brau99}. Thus, it suffices to treat the corresponding effects perturbatively, \ie we can assume the background to be flat at leading order.

If solely position space is curved, it is customary to construct Riemann normal coordinates $x^i$ and expand around a point $x_0$ \cite{Brewin09}. If both, position and momentum space, are curved, the corresponding curvature tensors (denoted $R_{ikjl}$ in position space and $S^{ikjl}$ in momentum space) can still be defined even though they can be position and momentum-dependent. Corrections to the metric are expected to be analogous to the Riemann-normal-coordinate expansion. In particular, mixing terms between positions and momenta are not expected at first order due to their twofold suppression. Therefore, we propose to expand around a point $Y_0=(x_0,p_0)$ in phase space. In particular, at the low energies expected when considering quantum-gravity effects in particle dynamics, $p_0$ is best chosen to be the origin of momentum space. 

Consequently, we can express the metric as
\begin{equation}
    g_{ij}\simeq\delta_{ij}-\frac{1}{3}\left(R_{ikjl}|_{Y_0}x^kx^l-S_{i~j}^{~k~l}|_{Y_0}p_kp_l\right),
\end{equation}
where indices are raised and lowered using the flat metric $\delta_{ij}.$ The sign difference between the correction terms compensates for the fact that $g_{ij}$ denotes the inverse of the metric in momentum space. As a result, the position representation of the operator $\hat{\mu}$ reads approximately
\begin{equation}
    \hat{\mu}\Psi\simeq\left[1-\frac{1}{12}\left(R_{kl}|_{Y_0}x^kx^l+S^{kl}|_{Y_0}\partial_k\partial_l\right)\right]\Psi,
\end{equation}
with the Ricci tensors in position and momentum space $R_{ij}$ and $S^{ij},$ respectively.

In normal coordinates, the classical geodesic distance from the origin is generally given as
\begin{equation}
	\sigma_p^2=g^{ij}p_ip_j.
\end{equation}
Therefore, the position representation of the free-particle Hamiltonian reads
\begin{align}
    \hat{H}_{\text{fp}}\Psi\simeq& \left(-\frac{\Delta_0}{2M}+\hat{H}^{(2)}_{\text{fp}}\right)\Psi,\\
    \hat{H}^{(2)}_{\text{fp}}\Psi     =&\frac{1}{6M}\Big[R^j_i|_{Y_0}x^i\partial_j-R^{i~j~}_{~k~l}|_{Y_0}x^kx^l\partial_i\partial_j\Big]\Psi,\label{hfp}
\end{align}
with the Laplacian in flat space $\Delta_0,$ and where we used the symmetries of the curvature tensors and absorbed a constant term into an unobservable redefinition of the ground-state energy.

\subsection{Time-independent perturbation theory}

A formalism dealing with non-singular perturbation theory additionally exerting an influence on the scalar product measure was introduced in  \cite{Dabrowski20}. Written in terms of general density-valued wave functions $\Psi,$ this reduces to an instant of ordinary perturbation theory. Under the assumption that $\braket{\Psi|\Psi}=\braket{\Psi^{(0)}|\Psi^{(0)}}=1,$ we then obtain
\begin{equation}
    \int\D^dx\text{Re}\left(\Psi^{*(2)}\Psi^{(0)}\right)=0.\label{wavedenspert}
\end{equation}
Say, a Hermitian operator $\op$ with discrete eigenvalues $\lambda_n$ and eigenstates $\Psi_n$ ($n$ can stand for several quantum numbers) is corrected by second-order contributions to the metric as $\op\simeq\op^{(0)}+\op^{(2)}.$ Then, Eq. \eqref{wavedenspert} implies for the corrections $\lambda^{(2)}_n$ to its eigenvalues $\lambda^{(0)}$ that
\begin{align}
    \lambda_n^{(2)}=\int\D^dx\Psi_n^{*(0)}\op^{(2)}\Psi_n^{(0)}.\label{eigvalpert}
\end{align}
In the given setup $\Psi^{(0)}=\psi^{(0)},$ which is why we can write
\begin{align}
    \lambda_n^{(2)}=\int\D^dx\psi_n^{*(0)}\op^{(2)}\psi_n^{(0)}.\label{eqn:defExpVal}
\end{align}
This latter equality is restricted to normal coordinates. In short, the modifications to the flat-space measure do not affect the eigenvalues at next-to-leading order. Note, however, that the measure is implicit to the kinetic term of the Hamiltonian given in Eq. \eqref{eqn:DefKinEnCurvCot}. Thus, it being nontrivial has an effect, \eg, on the eigenenergies of particles.
	
\subsection{Central potentials in three dimensions}

Central potentials approximate an interaction between the considered particle and an external object, which is massive enough to neglect backreaction on it. For example, the electron in a hydrogenic atom is usually modelled to experience a Coulomb-force emanating from the excessively more massive nucleus. The latter thus provides a preferred frame, where it is located at rest in the origin.

In ordinary quantum mechanics, Galilean invariance requires that interactions between particles depend solely on their respective distance. If only space is curved and in Riemann normal coordinates, this distance just reads $\sigma=r=\sqrt{x^ix^j\delta_{ij}}.$ This changes, however, when momentum space is allowed to become nontrivial, for example in deformations of Galilean relativity \cite{Bosso:2023nst}. In general, the geodesic distance from the origin, which in the given coordinate system coincides with $x_0,$ satisfies the differential equation 
\begin{equation}
    g^{ij}\partial_i\sigma\partial_j\sigma=1\label{geoddistdiff}
\end{equation} 
and the initial condition $\sigma|_{x_0}=0.$ In the perturbative case, we then expand $\sigma\simeq\sigma^{(0)}+\sigma^{(2)},$ with $\sigma^{(0)}=r.$ Furthermore, linearising Eq. \eqref{geoddistdiff}, we obtain the correction
\begin{align}
    \sigma^{(2)}=\frac{x^ix^j}{6r}S_{i~j}^{~k~l}p_kp_l .
\end{align}
Thus, we can expand any central potential as
\begin{equation}
    V(\sigma)=V(r)+\frac{V'(r)}{6}\frac{x^ix^j}{r}S_{i~j}^{~k~l}|_{Y_0}p_kp_l, \label{genpot}
\end{equation}
which, depending on the curvature tensor in momentum space can become anisotropic. 

As this potential is both position and momentum-dependent, promoting it to a Hermitian operator is an ambiguous task. The resulting operator ordering ambiguities have to be treated with care on a case-by-case basis, which will be done in the following examples.

%Taking into account all possible linear orderings, we obtain the correction to the potential operator
%\begin{align}
%    \hat{V}^{(2)}\Psi=&-\frac{\hbar^2}{12}\Big[V'\frac{x^ix^j}{r}S_{i~j}^{~k~l}\partial_k\partial_l+\frac{V'}{r}S^j_i x^i\partial_j\nonumber\\
%    &+\alpha S\frac{ V'}{r}+S_{ij}\frac{x^ix^j}{r^2}\left(\frac{\beta V'}{r}-\gamma V''\right)\Big]\Psi
%\end{align}
%where the parameters $\alpha,\beta,\gamma\in [0,1]$ parametrize the ambiguity which take the values $1/4,$ $1/3$ and $1/3$ respectively if all orderings are weighted equally. 

\subsubsection{Isotropic harmonic oscillator}

According to Eq. \eqref{genpot}, the potential describing an harmonic oscillator reads
\begin{equation}
    V(\sigma)\simeq\frac{1}{2}M\omega^2\left(r^2+\frac{1}{3}x^ix^jS_{i~j}^{~k~l}|_{Y_0}p_kp_l\right),
\end{equation}
with the oscillation frequency $\omega.$ Taking into account all possible operator orderings combining the four non-commuting contributions $\hat{x}^i,\hat{x}^j,\hat{p}_k,\hat{p}_l$, we obtain the correction to the corresponding operator
\begin{align}
    \hat{V}_{(2)}=&\frac{1}{6}M\omega^2\left(\hat{x}^i\hat{x}^jS_{i~j}^{~k~l}|_{Y_0}\hat{p}_k\hat{p}_l+i S^j_i|_{Y_0}\hat{x}^i\hat{p}_j\right),\label{harmoscpot}
\end{align}
where the dependence on the ordering is constant and can be removed by another redefinition of the ground-state energy. A close comparison of Eqs. \eqref{hfp} and \eqref{harmoscpot} makes apparent that the implications of curvature in both position and momentum space are analogous and Born reciprocity, a crucial feature of the harmonic oscillator mentioned in the introduction, is restored in adapted "natural units" ($M=\omega=1$).

The eigenstates and -values of the unperturbed Hamiltonian describing the harmonic oscillator $\hat{H}_{(0)}=\hat{H}_{\text{fp}}^{(0)}+\hat{V}_{(0)}$ read in the position representation and in spherical coordinates \cite{Messiah99}
\begin{align}
    \psi_{nlm}^{(0)}=&N_{nl}r^le^{-\frac{M\omega}{2}r^2}L_{\frac{n-l}{2}}^{l+1/2}\left(M\omega r^2\right)Y^l_{m}(\theta,\phi),\\
    N_{nl}=&\sqrt{\sqrt{\frac{M^3\omega^3}{\pi}}\frac{2^{\frac{n+3l}{2}+3}\left(\frac{n-l}{2}\right)!\left(\frac{M\omega}{2}\right)^l}{\left(2n+l+1\right)!!}},\\
    E^{(0)}_n=&\braket{\psi_{nlm}^{(0)}|\hat{H}_{(0)}\psi_{nlm}^{(0)}}=\omega\left(n+\frac{3}{2}\right),
\end{align}
with the generalized Laguerre polynomials $L_n^\alpha (x),$ the spherical harmonics $Y_{lm}(\theta ,\phi),$ and where we introduced the quantum numbers $n\geq 0$ (radial), $l=n\Mod{2},n\Mod{2}+2,\dots n$ and $m=-l,-l+1,\dots,l$ (angular). The curvature in both position and momentum space induces corrections, which can be calculated according to Eq. \eqref{eigvalpert} as
\begin{equation}
    E^{(2)}_{nlm}=\int\D^3x\Psi_{nlm}^{(0)}\hat{H}^{(2)}\Psi_{nlm}^{(0)}.
\end{equation}
These integrals can be solved analytically. Consequently, we obtain the corrections
\begin{align}
    E^{(2)}_{nlm}=\frac{\omega}{6}&\left[\left(l(l+1)-3m^2\right)\left(\frac{M\omega}{2}S_{zz}+\frac{R_{zz}}{2M\omega}\right)\left.+m^2\left(\frac{M\omega}{2}S+\frac{R}{2M\omega}\right)\right]\right|_{Y_0},\label{Ecorrharm}
\end{align}
with the Ricci scalars in position and momentum space $R$ and $S$, respectively.

In general, quantum harmonic oscillators are composite objects. As a reaction to experimental results in this context \cite{Pikovski12,Bawaj14}, it has been pointed out \cite{Quesne:2009vc,Tkachuk:2012gyq,Amelino-Camelia13,Kumar20,Bosso:2023aht} that quantum gravity effects as embodied by the GUP do not scale with powers of the Planck mass but its product with the number of fundamental constituents. The corresponding relative corrections at Planckian-per-constituent momentum space-curvature $S\sim (Nm_p)^{-2},$ with the effective number of constituents (elementary particles) contained in the oscillator $N,$ read for table top experiments
\begin{equation}
    \delta E_{nlm}\equiv\frac{E^{(2)}_{nlm}}{E^{(0)}_{nlm}}\sim\frac{M\omega}{N^2m_p^2}+\frac{E_{\text{surf}}^2}{M\omega},\label{harmoscres}
\end{equation}
with the energy scale corresponding to the spatial curvature acting on objects on the surface of the earth $E_{\text{surf}}=\sqrt{R}|_{\text{surf}}\sim 10^{-19}\text{eV}.$ Essentially, those effects are important in reciprocal regimes -- classical gravity modifies processes at small energies, \ie large distances and oscillation periods, while quantum gravity acts at high energies or small distances as expected. In particular, in the regime of high frequency and replacing $N=M/\bar{m}_c,$ with the average mass of the constituents $\bar{m}_c$, the corrections are of the form
\begin{equation}
    \delta E_{nlm}\sim\frac{\omega}{M}\frac{\bar{m}_c^2}{m_p^2}.
\end{equation}
This implies that it is most effective to use probes with high average constituent mass. Furthermore, the ratio $\omega/M$ favours microscopic oscillators, \ie such containing a small number of constituents. For the applications considered in the literature, \eg in  \cite{Pikovski12}, the relative corrections are minute once the number of constituents is taken into account in the way indicated here, which corresponds to the parameter $\alpha=1$ in  \cite{Kumar20}. For example, the setup in  \cite{Bawaj14} leads to relative corrections of $\delta E_{nlm}\sim 10^{-70}.$

More importantly, Eq. \eqref{harmoscres} implies that it is impossible to distinguish position from momentum space-curvature with a single harmonic oscillator of mass $M$ and frequency $\omega .$ More precisely, the corrections \eqref{harmoscres} are invariant under the exchange
\begin{equation}
    M\omega S_{ij}\longleftrightarrow\frac{R_{ij}}{M\omega},
\end{equation}
which is clearly reminiscent of T-duality in string theory \cite{Sathiapalan86}, connecting the low and high energy behaviour. This is not the first time T-duality has been encountered in the context of the curved cotangent bundle. In fact, it is manifest and as such one of the defining features of metastring theory and its underlying Born geometry \cite{Freidel13,Freidel14,Freidel15,Freidel17,Freidel18}. Thus, according to Eq. \eqref{harmoscres}, the harmonic oscillator maintains its Born reciprocity on arbitrary nontrivial backgrounds.

In a nutshell, under the assumption of a reciprocal Hamiltonian like the isotropic harmonic oscillator, we precisely obtain the behaviour Born was striving for, when trying to merge quantum theory with general relativity through the curved cotangent bundle.

\subsubsection{Coulomb potential}

The Coulomb potential approximates the dynamics of the electron wave function in a hydrogenic atom. The nucleus, possibly screened by further electrons, is modelled as a classical point with charge $Ze$ (where $Z\in \mathbb{N},$ and $e$ denotes the elementary charge) acting classically on the valence electron. In accordance with Eq. \eqref{genpot}, the classical coulomb potential reads to second order
\begin{equation}
    V(\sigma)\simeq\frac{Z\alpha}{r}\left(1-\frac{1}{6}S_{i~j}^{~k~l}|_{Y_0}\frac{x^ix^j}{r^2}p_kp_l\right),
\end{equation}
with the fine-structure constant $\alpha\simeq 1/137.$ The corresponding quantum operator can be found unambiguously under the assumption that the perturbative approach is consistent. All ordering-dependent corrections scale as $\hat{r}^{-3}.$ Thus, their expectation value with respect to the unperturbed eigenstates of the Hamiltonian diverges. Omitting them by choice of ordering is equivalent to renormalizing the problem, leading to the potential
\begin{equation}
    \hat{V}_{(2)}=-\frac{Z\alpha}{6\hat{r}^3}\left(S_{i~j}^{~k~l}|_{Y_0}\hat{x}^i\hat{x}^j\hat{p}_k\hat{p}_l+i S_i^j|_{Y_0}\hat{x}^i\hat{p}_j\right).
\end{equation}
At the unperturbed level, the hydrogenic atom has the eigenstates and -values \cite{Messiah99}
\begin{align}
    \psi_{nlm}^{(0)}=&N_{nl}\left(\frac{2nr}{a_0^*}\right)^le^{-\frac{nr}{a_0^*}}L_{n-l-1}^{2l+1}\left(\frac{2nr}{a_0^*}\right)Y^l_m(\theta,\phi),\\
    N_{nl}=&\sqrt{\left(\frac{2}{na_0^*}\right)^3\frac{(n-l-1)!}{2n(n+l)!}},\\
    E^{(0)}_n=&-\frac{Z^2\alpha^4M}{2n^2},
\end{align}
with the reduced Bohr radius $a_0^*=1/Z\alpha M$ and the quantum numbers $n>0,$ $l=0,1,\dots,n$ and $m=-l,-l=1,\dots,l.$

In accordance with Eq. \eqref{eqn:defExpVal}, the corrections to the energy levels read
\begin{equation}
	E^{(2)}_{nlm}=\int\D^3 x (\psi_{nlm}^{(0)})^*\hat{H}^{(2)}\psi_{nlm}^{(0)}
\end{equation}
On the one hand, if $l=0,$ we obtain
\begin{equation}
    E_{n00}^{(2)}=0.
\end{equation}
On the other hand, if $l\neq 0,$ the resulting corrections read
\begin{align}
    E^{(2)}_{nlm}=&\frac{M}{6}\left[\left(l(l+1)-3m^2\right)\left(C_{ln}M^2S_{zz}+\frac{R_{zz}}{2M^2}\right)\left.+m^2\left(C_{ln}M^2S+\frac{R}{2M^2}\right)\right]\right|_{Y_0},\label{coulres}
\end{align}
where we defined the function of the quantum numbers
\begin{equation}
    C_{nl}=-\frac{Z^2\alpha^4}{n^3l(l+1)(2l+1)}.
\end{equation}
Again, the spatial curvature acts most strongly on large length scales (small $M$) while the curvature in momentum space is most effective for small distances (large $M$). Neglecting the contribution of the spatial curvature, the effect is largest for $n=2,$ $l=1$ and $m=0.$ It strongly decays towards higher values of $n$ and $l$ and vanishes for $l=0.$ At Planckian curvature scales $S_{ij}\sim m_p^2,$ \ie assuming that $M$ reflects the mass of an elementary particle, the relative correction to the energy is of the order
\begin{equation}
    \delta E_{nlm}\sim\frac{M^2}{m_p^2}.
\end{equation}
For the hydrogen atom, where $M$ denotes the mass of the electron, this corresponds to a value $\delta E_{nlm}\sim 10^{-44}$ as had been reported before in the context of the GUP \cite{Bouaziz10,AntonacciOakes13,Brau99}. This result can be improved upon by applying it to more massive charged particles such as the $W^{\pm},$ yielding $\delta E_{nlm}\sim 10^{-35}.$  

Over all, the corrections to the bound-state energies of the hydrogenic atom \eqref{coulres} and  the harmonic oscillator are similar \eqref{Ecorrharm}. For fixed $n$ and $l,$ there continues to be an invariance under the transformation
\begin{equation}
    C_{nl}M^2S_{ij}\longleftrightarrow \frac{R_{ij}}{2M^2},
\end{equation}
which, considering the inverse scaling with the squared mass, again has a taste of T-duality. However, taking into account states of distinct $l$ and/ or $n,$ this symmetry is broken in general -- a hydrogenic atom could clearly be used as a means of discriminating between curvature in position and momentum space. This was to be expected because the underlying Hamiltonian explicitly breaks Born reciprocity, thereby also breaking the T-duality-like invariance.

\section{Conclusion}\label{sec:concl}

Recent developments at the intersection of the minimal-length paradigm and nontrivial momentum space necessitate the creation of a formalism, capable of describing quantum mechanics on backgrounds with curvature in position and momentum space. In particular, this approach is required to investigate the position representation of quantum mechanics on a nontrivial momentum space, which was recently \cite{Wagner21a} found to be dual to GUP-deformed quantum theory. Furthermore, this formalism sets the stage to find valid representations of the GEUP, a task which had proven illusive thus far.

The present paper marks a first step towards such a description, taking a rather intuitive route. In particular, we have promoted the Hilbert-space measure, derived from the metric \'a la deWitt \cite{DeWitt52}, to an operator. Subsequently, we have split it into two pieces and merged those symmetrically with the wave functions entering the scalar product. We thus obtained scalar-density valued wave functions by analogy with the geometric quantization program \cite{Woodhouse97}. Consequently, under the assumption that they comply with the geometrical modifications to derivatives derived from the nonlinear connection on the curved cotangent bundle, we have defined the momentum operator in the position representation by analogy with its undeformed counterpart. Furthermore, we have found a representation of the Hamiltonian of a free particle and, analogously, the geodesic distance to the origin.

This has made it possible to investigate a generic metric which we have expanded in position and momentum space by analogy with Riemann normal coordinates. After a short discussion of perturbation theory in this context, we have applied the formalism to two central potentials, the harmonic oscillator and the hydrogenic atom. As a result, we have obtained analytical corrections to the eigenvalues of the Hamiltonian in terms of the Ricci scalars and tensors in position and momentum space. Remarkably, the isotropic harmonic oscillator retains its symmetry between positions and momenta, thus, in principle, making it impossible to distinguish between curvature in position and momentum space. Consequently, we have achieved an instantiation of Born reciprocity in quantum mechanics on the curved cotangent bundle - exactly in the way originally intended by Born \cite{Born38,Born49}. This property is accompanied by a T-duality-like behaviour of the ensuing relative corrections to the energies of stationary states. Thus, the findings presented in this paper corroborate the relation between T-duality, Born reciprocity and the curved cotangent bundle manifest in metastring theory and Born geometries \cite{Freidel13,Freidel14,Freidel15,Freidel17,Freidel18}.

These encouraging results will make it possible to investigate the position representation of theories of curved momentum space dual to noncommutative geometries and GUPs or GEUPs, \ie instances of quantum spaces. This analysis will be the subject of future research.

\section*{Acknowledgements}

The author thanks Luciano Petruzziello and Mariusz D\k{a}browski for insightful discussions. Furthermore, he is indebted to the anonymous reviewer who provided valuable comments which helped improve the paper. His work was supported by the Polish National Research and Development Center (NCBR) project ''UNIWERSYTET 2.0. --  STREFA KARIERY'', POWR.03.05.00-00-Z064/17-00 (2018-2022). Moreover, the author would like to acknowledge the contribution of the COST Action CA18108.

\bibliographystyle{utphys} % We choose the "plain" reference style
\bibliography{ref}

\end{document}